\documentclass[pdflatex,sn-mathphys-num]{sn-jnl}% Math and Physical Sciences Numbered Reference Style
\usepackage{graphicx}%
\usepackage{multirow}%
\usepackage{amsmath,amssymb,amsfonts}%
\usepackage{amsthm}%
\usepackage{mathrsfs}%
\usepackage[title]{appendix}%
\usepackage{xcolor}%
\usepackage{textcomp}%
\usepackage{manyfoot}%
\usepackage{booktabs}%
\usepackage{algorithm}%
\usepackage{algorithmicx}%
\usepackage{algpseudocode}%
\usepackage{listings}%
\usepackage{natbib}
\theoremstyle{thmstyleone}%
\theoremstyle{thmstyletwo}%

\theoremstyle{thmstylethree}%

\begin{document}

\title[Agentic Artificial Intelligence in Finance: A Comprehensive Survey]{Agentic Artificial Intelligence in Finance: A Comprehensive Survey}


\author*[1]{\fnm{Irene} \sur{Aldridge}}\email{irene.aldridge@gmail.com}
 
\author[1]{\fnm{Jolie} \sur{An}}\email{xa39@cornell.edu}
\equalcont{These authors contributed equally to this work.}

\author[1]{\fnm{Riley} \sur{Burke}}\email{rgb234@cornell.edu}
\equalcont{These authors contributed equally to this work.}

\author[1]{\fnm{Michael} \sur{Cao}}\email{yc849@cornell.edu}
\equalcont{These authors contributed equally to this work.}

\author[1]{\fnm{Chia-Yi} \sur{Chien}}\email{cc2889@cornell.edu}
\equalcont{These authors contributed equally to this work.}

\author[1]{\fnm{Kexin} \sur{Deng}}\email{kd537@cornell.edu}
\equalcont{These authors contributed equally to this work.}

\author[1]{\fnm{Ruipeng} \sur{Deng}}\email{rd639@cornell.edu}
\equalcont{These authors contributed equally to this work.}

\author[1]{\fnm{Yichen} \sur{Gao}}\email{yg635@cornell.edu}
\equalcont{These authors contributed equally to this work.}

\author[1]{\fnm{Olivia} \sur{Guo}}\email{qg77@cornell.edu}
\equalcont{These authors contributed equally to this work.}

\author[1]{\fnm{Shunran} \sur{He}}\email{sh2662@cornell.edu}
\equalcont{These authors contributed equally to this work.}

\author[1]{\fnm{Zheng} \sur{Li}}\email{zzl2@cornell.edu}
\equalcont{These authors contributed equally to this work.}

\author[1]{\fnm{George} \sur{Lin}}\email{jl2969@cornell.edu}
\equalcont{These authors contributed equally to this work.}

\author[1]{\fnm{Weihang} \sur{Lin}}\email{wl769@cornell.edu}
\equalcont{These authors contributed equally to this work.}

\author[1]{\fnm{Percy} \sur{Lyu}}\email{fl466@cornell.edu}
\equalcont{These authors contributed equally to this work.}

\author[1]{\fnm{Alex} \sur{Ng}}\email{kn438@cornell.edu}
\equalcont{These authors contributed equally to this work.}

\author[1]{\fnm{Qi} \sur{Wang}}\email{qw297@cornell.edu}
\equalcont{These authors contributed equally to this work.}

\author[1]{\fnm{Hanxi} \sur{Xiao}}\email{hx299@cornell.edu}
\equalcont{These authors contributed equally to this work.}

\author[1]{\fnm{Dora} \sur{Xu}}\email{dx33@cornell.edu}
\equalcont{These authors contributed equally to this work.}

\author[1]{\fnm{Yuanyuan} \sur{Xue}}\email{yx665@cornell.edu}
\equalcont{These authors contributed equally to this work.}

\author[1]{\fnm{Sheng} \sur{Zhang}}\email{sz696@cornell.edu}
\equalcont{These authors contributed equally to this work.}

\author[1]{\fnm{Sirui} \sur{Zhang}}\email{sz694@cornell.edu}
\equalcont{These authors contributed equally to this work.}

\author[1]{\fnm{Yun} \sur{Zhang}}\email{yz2847@cornell.edu}
\equalcont{These authors contributed equally to this work.}

\author[1]{\fnm{Sirui} \sur{Zhao}}\email{sz695@cornell.edu}
\equalcont{These authors contributed equally to this work.}

\author[1]{\fnm{Xiaolong} \sur{Zhao}}\email{xz935@cornell.edu}
\equalcont{These authors contributed equally to this work.}

\author[1]{\fnm{Yihan} \sur{Zhao}}\email{yz3337@cornell.edu}
\equalcont{These authors contributed equally to this work.}

\author[1]{\fnm{Waner} \sur{Zheng}}\email{wz462@cornell.edu}
\equalcont{These authors contributed equally to this work.}

\affil*[1]{\orgdiv{ORIE, Financial Engineering}, \orgname{Cornell University}, \orgaddress{\street{214 Frank L. Porter Hall}, \city{Ithaca}, \state{NY}, \postcode{14863}, \country{USA}}}


\abstract{The emergence of agentic artificial intelligence (AI) represents a fundamental transformation in financial markets, characterized by autonomous systems capable of reasoning, planning, and adaptive decision-making with minimal human intervention. This comprehensive survey synthesizes recent advances in agentic AI across multiple dimensions of financial operations, including system architecture, market applications, regulatory frameworks, and systemic implications. We examine how agentic AI differs from traditional algorithmic trading and generative AI through its capacity for goal-oriented autonomy, continuous learning, and multi-agent coordination. Our analysis shows that while agentic AI offers substantial potential for enhanced market efficiency, liquidity provision, and risk management, it also introduces novel challenges related to market stability, regulatory compliance, interpretability, and systemic risk. Through a systematic review of foundational research, technical architectures, market applications, and governance frameworks, this survey provides scholars and practitioners with a structured understanding of how agentic AI is reshaping financial markets and identifies critical research directions for ensuring that these systems enhance both operational efficiency and market resilience.
}




\maketitle

\section{Introduction}\label{sec1}

Financial markets are experiencing a profound transformation as artificial intelligence evolves from narrow, task-specific automation toward agentic systems capable of autonomous reasoning, strategic planning, and adaptive coordination. Over the past decade, AI has progressed from rudimentary rule-based systems performing discrete functions to sophisticated multi-agent architectures that plan, learn, and act semi-autonomously under human oversight. This shift from automation to agency represents more than a technological upgrade; it signals a fundamental restructuring of how financial decisions are made, risks are managed, and markets function.

Agentic AI systems differ from their predecessors in three critical dimensions. First, they possess goal-oriented autonomy, allowing them to set intermediate objectives and decompose complex tasks without continuous human instruction. Second, they employ contextual reasoning, integrating perception, memory, and planning to adapt strategies based on evolving market conditions. Third, they engage in multi-agent collaboration, coordinating with other intelligent systems through structured communication protocols to achieve collective outcomes. These capabilities enable financial institutions to move beyond static algorithmic execution toward dynamic, adaptive decision ecosystems where machines handle scale and pattern recognition while humans provide strategic oversight and ethical judgment.

The operational and strategic implications of agentic AI extend across multiple domains of financial markets. In trading, multi-agent reinforcement learning systems now replicate and enhance institutional behaviors through coordinated execution strategies. In portfolio management, agents dynamically rebalance allocations based on regime detection and predictive state representations. In risk management, autonomous systems perform real-time stress tests and scenario analysis at scales previously infeasible. In compliance, agentic architectures automate regulatory monitoring while maintaining audit trails for accountability. In all of these applications, agentic AI is redefining how expertise, control, and accountability are distributed within financial institutions.

However, this technological evolution introduces significant challenges that require the attention of researchers and practitioners. The opacity of multi-agent decision processes raises questions about interpretability and regulatory compliance. The adaptive nature of learning algorithms creates non-stationarity that complicates traditional validation and risk assessment frameworks. The interconnection of autonomous systems across institutions amplifies the potential for correlated behaviors and systemic instability. The concentration of computational power and data infrastructure risks creating competitive asymmetries and market dependencies. Addressing these challenges requires not only technical innovation but also fundamental rethinking of governance structures, evaluation methodologies, and market design principles.

This survey provides a comprehensive synthesis of the emerging literature on agentic AI in finance, organized around four core research streams. First, we examine the foundational concepts that distinguish agentic systems from classical and generative AI, establishing the definitional clarity essential for rigorous analysis. Second, we analyze technical architectures that enable autonomous decision-making, including cognitive frameworks, memory systems, reasoning modules, and communication protocols. Third, we review applications across key financial domains, from trading and portfolio optimization to risk management and regulatory compliance. Fourth, we assess systemic implications for market structure, including impacts on liquidity, price discovery, resilience, and competitive dynamics. Finally, we synthesize current regulatory approaches and governance frameworks, identifying critical gaps between technological capabilities and oversight mechanisms. 

\subsection{Motivation and Scope}

The motivation for this survey stems from the recognition that agentic AI represents a discontinuous shift in the way financial markets operate, demanding fresh theoretical frameworks and empirical methodologies. While substantial research exists on algorithmic trading, machine learning in finance, and market microstructure, the literature on truly autonomous, multi-agent financial systems remains fragmented across computer science, economics, and operations management. This survey bridges these disciplinary perspectives, providing operations management scholars with a structured understanding of both the technical foundations and the practical implications of agentic AI. We consider all papers in the realm of "agentic AI", "finance", and "reinforcement learning" Scopus-indexed as of December 23, 2025, as well as additional preprints not on the Scopus radar (such as those published on SSRN).  

Our scope encompasses several interconnected research areas. We review agent design patterns and architectures that define how autonomous systems organize internal components for perception, reasoning, and action. We examine multi-agent reinforcement learning frameworks that enable coordination and competitive strategy development. We analyze how agentic systems affect the properties of the market microstructure such as depth, resilience, and price discovery. We investigate regulatory frameworks and governance mechanisms designed to maintain accountability and stability as autonomy increases. Throughout, we emphasize the tension between individual optimization and collective stability, a central concern for operations management in complex adaptive systems.

This survey deliberately focuses on systems that exhibit genuine agency—autonomy in goal formulation, adaptive learning, and strategic coordination—rather than merely automated or predictive systems. We exclude traditional algorithmic trading that executes fixed strategies, basic machine learning models that generate predictions without autonomous action, and rule-based compliance systems that lack adaptive capabilities. By maintaining this focus, we isolate the unique challenges and opportunities presented by the evolution toward truly autonomous financial agents.

\subsection{Contribution and Organization}

This survey makes several contributions to the literature. First, it provides definitional clarity by distinguishing agentic AI from related concepts and establishing a taxonomy of agency dimensions. Second, it synthesizes technical architectures from computer science into frameworks accessible to operations researchers, highlighting decision structures relevant for system design and evaluation. Third, it systematically reviews empirical evidence on market impacts, identifying both benefits and risks through the lens of market microstructure theory. Fourth, it analyzes regulatory and governance challenges through the perspective of accountability mechanisms and mechanism design. Fifth, it identifies persistent gaps in evaluation methodologies, interpretability techniques, and systemic risk assessment—bridging academic research and practical deployment.

The remainder of this survey is organized as follows. Section \ref{sec2} establishes the foundational concepts that define agentic AI and its distinguishing characteristics. Section \ref{sec3} examines technical architectures, covering cognitive frameworks, memory systems, planning mechanisms, and multi-agent communication. Section \ref{sec4} reviews applications in trading, portfolio management, and risk operations. Section \ref{sec5} analyzes the impacts of the market structure, including effects on liquidity, resilience, and competitive dynamics. Section \ref{sec6} addresses regulatory landscapes and governance frameworks. Section \ref{sec7} synthesizes the research challenges and future directions. Section \ref{sec8} concludes. 

\section{Foundational Concepts and Definitions of Agentic AI}\label{sec2}

The concept of agentic artificial intelligence has emerged as a defining paradigm in AI research, representing a fundamental evolution beyond simple automation and static prediction models. Understanding what constitutes "agency" in artificial systems is essential for analyzing how these systems differ from traditional approaches and what unique challenges and opportunities they present for financial markets.

\subsection{Defining Agentic AI}

The notion of "agentic" intelligence originates from the broader concept of agency, which refers to an entity's capacity to act independently and make choices. Within AI, agency involves a system's ability to set goals, plan actions, learn from feedback, and execute decisions without continuous external input. Nisa et al. (2025) define agentic AI as "AI systems capable of autonomous reasoning and goal-driven action in dynamic environments," emphasizing that such systems integrate perception, planning, and execution to operate continuously on open-ended tasks.

A critical distinction emerges between reactive AI systems that passively respond to predefined prompts and proactive agentic systems that actively formulate goals and decompose tasks. The taxonomy proposed in recent research distinguishes agentic AI from traditional AI agents along a spectrum of capability—from reactive agents that respond merely to stimuli to self-aware agents capable of metacognitive reasoning. The key differentiation is the degree of autonomy and goal formulation: agentic AI systems not only execute actions, but also decide what to do next based on evolving contexts.

Recent conceptual work frames agentic AI as a natural progression from generative models. Whereas generative AI produces content based on prompts, agentic AI engages in continuous reasoning and decision-making loops—decomposing tasks, selecting tools, and assessing intermediate outcomes. This progression from generation to agency represents a shift from content creation to autonomous problem-solving and strategic action.

\subsection{Core Characteristics of Agentic Systems}

Five core characteristics distinguish agentic AI systems from traditional automation and predictive models:
\paragraph{Autonomy}
Autonomy is the defining characteristic of agentic AI, referring to a system's ability to make independent decisions without step-by-step human guidance. This autonomy is both behavioral—the ability to execute tasks—and cognitive—the ability to set and justify goals. From a governance perspective, autonomy must coexist with accountability: autonomous systems should remain interpretable and ethically aligned. This duality of freedom and responsibility defines the modern debate surrounding agency in AI and is particularly acute in financial applications where autonomous decisions carry significant monetary and regulatory consequences.

Autonomy is often achieved through reinforcement learning (RL). Reinforcement learning reframes trading as sequential decision-making under uncertainty, directly optimizing risk-adjusted returns by interacting with an environment (Pippas et al., 2024). An agent's objective often follows the standard formulation:
\begin{equation}
    max_\pi \mathbb{E}\left[\sum_{t=0}^T \gamma^t R_t\right]
\end{equation}

where $R_t$ denotes the reward at time $t$, $\gamma$ is the discount factor, and $\pi$ represents the state mapping policy with actions. In trading contexts, rewards typically incorporate volatility-scaled returns:
\begin{equation}
 R_t = \frac{r_t}{\sigma_t}   
\end{equation}
ensuring that exposure decreases when volatility rises (Zhang et al., 2020).
Recent surveys synthesize RL applications across portfolio allocation, execution, and market making, noting advances in policy optimization, exploration under non-stationarity, and reward shaping (Pippas et al., 2024). RL connects naturally to agent-based perspectives where multiple heterogeneous agents co-evolve and interact.

\paragraph{Reasoning and Planning}
Reasoning enables agentic AI systems to infer, generalize, and make context-aware decisions, while planning allows them to structure multistep processes. Architectures such as ReAct (Reason + Act) and Reflection allow agents to deliberate before execution, combining internal reasoning with external action. This reasoning-based planning distinguishes agentic systems from reactive large language models that perform one-shot prompt completion. Instead, these systems employ internal memory and feedback loops to continuously refine actions, mirroring human cognitive cycles of hypothesis formation, action execution, and outcome evaluation.

\paragraph{Adaptability and Self-Improvement}
Agentic AI must operate in non-stationary, real-world environments where conditions change unpredictably. Adaptability refers to a system's capacity to learn from environmental feedback and modify internal models accordingly. This capability is essential for achieving long-term competence—for example, adjusting trading strategies as market regimes shift or re-prioritizing risk management protocols based on emerging threats. Adaptability is tied to model-based reasoning, enabling agents to anticipate outcomes and revise behaviors proactively rather than merely reacting to observed patterns. This self-improvement capacity distinguishes agentic AI from static systems that rely solely on pre-trained parameters.

\paragraph{Multi-Agent Collaboration and Tool Use}
Modern agentic AI systems often function within ecosystems rather than in isolation. Recent architectural patterns introduce coordination layers that manage multiple agents and external tools, allowing communication, delegation of subtasks, and integration with databases and APIs. This distributed coordination mirrors human organizational structures, facilitating complex task decomposition and cooperative problem-solving. For instance, one agent may handle analytical reasoning while another executes trades, with a supervising agent coordinating their activities and resolving conflicts. Such multi-agent architectures enhance robustness and scalability by distributing cognitive and computational load across specialized components.
\paragraph{Long-Term Goals and Multistep Workflows}
Unlike reactive AI that responds to immediate input, agentic AI systems pursue long-term objectives through multistep reasoning chains. This temporal dimension introduces new challenges in alignment and predictability. long-term autonomy requires persistent memory, evaluation metrics, and feedback loops to maintain goal coherence over long time horizons. Architecturally, planning-reflection-execution loops exemplify this trait, enabling agents to maintain continuity across extended workflows such as continuous trading, autonomous research, or multistage data analysis. The ability to maintain goal-directed behavior over time is fundamental to genuine agency.

\subsection{Comparison with Classical and Generative AI}
The evolution from classical AI to agentic AI can be understood as a progressive shift in control, cognition, and continuity. Classical AI systems are human-driven and rely on explicit programming and symbolic reasoning for specific tasks. They excel at well-defined problems with clear rules, but struggle with ambiguity and adaptation. Generative AI represents an intermediate stage—it is prompt-driven, capable of producing creative output through learned statistical patterns, yet still dependent on human input for direction and task specification.

In contrast, agentic AI is self-driven: it integrates reflective reasoning, long-term planning, and adaptive decision-making to pursue goals continuously and autonomously. Although generative AI excels at creating novel outputs, it lacks sustained reasoning or goal awareness beyond the immediate prompt. Agentic AI bridges this gap by embedding generative capabilities within reasoning frameworks, using generative models as tools rather than endpoints. This shift represents a transformation from stochastic creativity to structured cognition, where the system not only generates responses, but also evaluates their appropriateness, adjusts strategies based on outcomes, and maintains coherent goal-directed behavior over extended interactions.

For financial applications, this distinction is critical. A classical algorithmic trading system executes predetermined strategies based on fixed rules. A generative AI system might predict price movements or generate investment reports based on prompts. However, an agentic trading system would continuously monitor markets, adapt its strategy based on regime changes, coordinate with other trading agents to optimize execution, and maintain risk controls autonomously—all while pursuing overarching portfolio objectives specified by human managers. This level of integration and autonomy fundamentally changes the nature of human-machine collaboration in financial operations.

\section{Technical Architecture and System Design}\label{sec3}

The deployment of agentic AI in financial markets requires sophisticated technical architectures that support autonomous reasoning, adaptive learning, and coordinated decision-making. This section examines the core architectural components that enable agentic systems to function effectively in complex financial environments, drawing on recent advances in cognitive architectures, memory systems, planning frameworks, and communication protocols.

\subsection{Cognitive Architectures for Financial Agents}

Beyond generating predictions or numeric scores, modern financial agents built on foundation models are designed to handle complex, multimodal inputs—such as text, prices, and structured financial indicators—and reason through the broader economic context before acting. Cognitive architecture refers to how an agent organizes its internal components for perception, memory, reasoning, and decision-making. It also describes how different roles or modules interact within a single agent or across multiple agents.

Zhu et al. (2023) provide a comprehensive foundation through their Agent Design Pattern Catalogue, which organizes over thirty design patterns commonly used in foundation-model-based agents. The patterns are divided into several functional categories: Prompting, Tool Use, Memory, Planning and Reasoning, Self-Reflection, and Multi-Agent Collaboration. Each pattern describes a specific way to organize the internal modules of an agent. For example, the ReAct pattern combines reasoning and acting in a loop, where the agent thinks about the next step before calling a tool or API. The Reflection pattern adds a self-evaluation stage, allowing the agent to review its previous outputs and improve them.

Wang et al. (2024) expand on these ideas through their FS-ReasoningAgent framework, which separates the reasoning process into two distinct paths: a fact reasoning path and a subjectivity reasoning path. The fact reasoning path handles objective information such as blockchain data, technical indicators, and factual news, while the subjectivity reasoning path focuses on investor opinions, social sentiment, and market mood. Each path is controlled by its own specialized sub-agent, and their outputs are combined by a trade decision agent. A reflection agent monitors the final results and adjusts how much weight to give each reasoning path in future trades. This multi-role setup allows the system to balance data-driven logic and sentiment-driven reasoning.

Yu et al. (2023) introduce a different structural approach with their FINMEM framework, built around memory and personality. The system has three main modules: a profiling module that defines the agent's risk tolerance and trading style, a memory module divided into short-term working memory and three long-term layers (shallow, intermediate, and deep) that store events with different levels of importance, and a decision module that retrieves key memories and combines them with current market state. By connecting memory retrieval, reflection, and personality in one loop, FINMEM produces trading behavior that is more stable and interpretable over time.

\subsection{Memory Systems and Knowledge Representation}
Financial agents rely on memory systems to keep track of information over time, learn from past events, and use that knowledge to guide future actions. In traditional trading algorithms, information is often processed in short batches without long-term awareness. In contrast, agent-based architectures introduce structured memory systems that allow agents to recall, summarize, and reuse context over extended periods, improving both performance and interpretability.

The FINMEM framework's memory module provides a clear example of structured memory design. The working memory acts as an immediate workspace, where recent data such as daily prices or market news are summarized before trading decisions. long-term memory is organized into three layers—shallow, intermediate, and deep—each corresponding to different time sensitivities. Daily news is deposited in the shallow layer, quarterly reports are in the intermediate layer, and annual reports or long-term performance summaries are in the deep layer. Information flows between layers according to importance and recency, allowing significant events to be retained longer. This layered structure helps agents remember both short-term trends and long-term fundamentals.

Wang et al. (2024) introduce a different approach through Sequential Knowledge-Guided Prompting (SKGP) in their LLMFactor framework. Rather than explicit memory storage, the system uses prompt-chaining, where language models generate intermediate knowledge representations before making predictions. The process builds background knowledge on company relationships, extracts factors from financial news, and combines these with textual representations of past price data. Although the model does not store memory across time in the traditional sense, its prompt-chaining structure functions as a temporary context-sensitive memory that enables interpretable reasoning aligned with human understanding of financial events.

Another widely adopted mechanism is Retrieval-Augmented Generation (RAG), which combines an external document retriever with a generative model. This allows agents to dynamically access relevant information at query time rather than relying solely on internal memory. In financial applications, RAG might retrieve specific filings, earnings reports, or market news to provide context for ongoing analysis. Although RAG does not maintain persistent memory of past reasoning steps, it acts as an on-demand memory extension that improves factual grounding and reduces hallucinations during reasoning.

\subsection{Planning and Reasoning Modules}
The planning and reasoning modules determine how a financial agent interprets information, makes intermediate decisions, and takes final actions. In traditional quantitative systems, reasoning is often embedded implicitly in statistical models or reinforcement learning policies. In contrast, LLM-based agents make reasoning an explicit part of the architecture through structured prompts, multi-step logic, or role-based coordination.

The FS-ReasoningAgent framework (Wang et al., 2024) addresses reasoning complexity by separating it into coordinated pathways for objective and subjective analysis. Each path is managed by a dedicated sub-agent, with outputs merged by a trade decision agent that weighs both factual and subjective evidence. A reflection agent monitors the results and adjusts the influence of each path over time, allowing the system to learn which reasoning pattern performs better under different market regimes. This structure makes the reasoning process more interpretable and reduces overreliance on any single information source.

The LLMFactor framework demonstrates sequential reasoning through its structured prompting approach. The agent first constructs background knowledge of company relationships, then identifies major factors from news, and finally integrates this with historical price data to make predictions. Each step acts as an intermediate reasoning layer that feeds into the next, acting as a form of planning where the model decides what to recall, analyze, and synthesize. Compared with end-to-end prediction models, this modular reasoning improves transparency and allows users to trace how conclusions are reached. 

Han et al. (2024) provide empirical evidence of how reasoning structures affect financial analysis through multi-agent collaboration. Their study compares single-agent, dual-agent, and triple-agent configurations across fundamental, sentiment, and risk analysis tasks. The results show that the effectiveness of reasoning depends on the complexity of the task and the communication structure. Single agents perform best in simple, data-driven tasks where consistent reasoning is valuable, while multi-agent setups with distinct roles for data collection, evaluation, and synthesis outperform in complex, ambiguous tasks like risk analysis. These findings highlight how reasoning structures and planning hierarchies directly influence agent performance and decision quality.

Zhu et al. (2023) expand reasoning architectures through patterns such as Chain-of-Thought, where agents produce step-by-step intermediate reasoning before action; ReAct, which alternates reasoning with tool use; and Simulation, where agents mentally explore multiple hypothetical outcomes before acting. These patterns are particularly relevant for financial agents that must evaluate multiple strategies under uncertainty. For example, simulation-style reasoning could compare several trading actions under different macroeconomic scenarios before committing, making decision-making both more interpretable and more testable against market data.

\subsection{Communication Protocols Between Agents}
As financial agents evolve from isolated systems to collaborative frameworks, communication protocols become critical to ensure that information is shared effectively and decisions remain coordinated. Communication refers to the structured exchange of messages, reasoning outputs, and intermediate analyzes between agents. Well-designed protocols allow agents to combine different analytical perspectives while maintaining stability and preventing conflicts.

Han et al. (2024) provide a systematic analysis of communication strategies in financial research. Their framework compares horizontal, vertical, and hybrid collaboration structures. In horizontal setups, all agents communicate equally, sharing reasoning steps and analyzes in open discussion. This structure performs best for straightforward analytical tasks where pooling independent insights increases completeness. Hybrid structures maintain a leader agent while allowing open discussion among subordinates, balancing autonomy and control. Vertical communication, where a single leader assigns and aggregates work, outperforms others in complex analytical tasks like risk assessment. The hierarchical design allows specialized subordinate agents to focus on component tasks while the leader consolidates results into unified decisions.

Fang et al. (2023) advance communication through their Intention-Aware Multi-Agent Communication (IMAC) framework. Instead of broadcasting raw observations, each trading agent communicates intentions—such as planned trade volume and direction—to other agents in multiple rounds. This allows agents to predict peers' actions and adjust strategies before execution. The IMAC model employs reinforcement learning with action-value attribution loss that measures how each agent's communication contributes to team reward. In simulated trading environments, this design leads to more efficient coordination, reduced resource conflicts, and improved portfolio performance. The multi-round protocol is especially relevant for systems involving constrained resources like capital allocation or liquidity management, where agents must negotiate rather than act independently.

Zhu et al. (2023) describe communication from an architectural perspective through reusable design templates including the Broadcasting Pattern, where agents share updates with all peers; the Negotiation Pattern, which enables cooperative problem-solving; and the Shared Workspace Pattern, which acts as a memory buffer accessible to multiple agents. These structures provide flexible building blocks for collaborative financial systems. When applied to trading or portfolio optimization, these patterns help define whether communication is synchronous (message passing in real time) or asynchronous (agents updating a shared knowledge base over time).

\subsection{Technical Infrastructure and Risk Management}
The deployment of agentic AI systems requires a robust technical infrastructure that supports stress testing and automated scenario analysis. Tirulo et al. (2025) provide a systematic framework distinguishing agentic intelligence from classic automation through three key features: autonomy of independent decisions, flexibility when dealing with changing environments, and decision-making initiative exceeding predefined responses.

Their proposed three-layer technology stack includes an Infrastructure Layer with edge and cloud computing for distributed processing, an Intelligence Layer with cognitive architectures and multi-agent system models, and a Decision and Learning Layer based on large language models and reinforcement learning algorithms. This architecture provides a blueprint for building automated stress testing systems capable of operating at various scales and complexity levels. The framework emphasizes world models—internal simulations agents use to envision outcomes without real-world experimentation—allowing agents to simulate thousands of scenarios mentally before implementation.

Bahrpeyma and Reichelt (2022) demonstrate multi-agent reinforcement learning (MARL) in dynamic environments through analysis of smart factory applications, which parallel financial trading in their uncertainty and coordination requirements. Their research shows that simulation-trained MARL-based systems must learn robust strategies through adversarial training and environmental perturbations. The distributed agent architecture proves to be more suitable than centralized control for handling unexpected system states, where individual agents respond locally to failures while coordinating globally. This approach directly informs stress testing methodologies for financial systems.

Lohn et al. (2023) provide a development pipeline for autonomous systems consisting of distinct testing stages: simulated laboratory tests in controlled ranges, testing bed validation against known scenarios, red-team testing with human opponents, and partial operational deployment. This progressive methodology serves as a template for risk management in autonomous financial systems. Their emphasis on cyber ranges—virtualized environments that enable automated red-teaming—illustrates how AI agents can explore defensive systems at machine speed, creating thousands of attack scenarios that are impractical for human testers. The continuous testing model they propose transitions from discrete phases to ongoing stress evaluation throughout operational life.

Lombardo et al. (2022) demonstrate fine-grained agent-based modeling for large-scale scenario analysis, showing how automated policy combinations can be evaluated to find effective interventions across uncertainty dimensions. Their sensitivity analysis identifying parameters with greatest impact provides a principled approach to prioritizing stress test scenarios. Applied to financial systems, this methodology enables automated scenario generation exploring environmental conditions, system failures, and adversarial behavior through combinatorial approaches, focusing on critical edge cases where small parameter changes cause large behavioral shifts.

\section{Applications in Financial Markets}\label{sec4}

Agentic AI systems are being deployed across multiple domains of financial markets, from high-frequency trading and portfolio management to risk assessment and regulatory compliance. This section examines key application areas where autonomous agents are transforming traditional financial operations, highlighting both the technical innovations and practical challenges encountered in real-world deployments.

\subsection{Multi-Agent Trading and Game Theory}

Modern financial markets can be viewed as multi-agent systems where autonomous participants—high-frequency algorithms, market makers, and institutional investors—interact strategically to maximize returns under uncertainty. Game theory provides the mathematical foundation for analyzing such strategic interactions, while multi-agent reinforcement learning adds an adaptive, data-driven layer allowing agents to learn policies from experience.

Classical game theory models rational agents making interdependent decisions, with the Nash equilibrium representing strategy profiles where no participant can improve utility by unilaterally changing strategy. In financial markets, this maps to trading equilibria where no trader can improve risk-adjusted returns given others' positions and expectations. However, traditional game theory assumes agents have perfect rationality and complete information—assumptions often violated in real markets. Multi-agent reinforcement learning relaxes these assumptions by allowing agents to learn from limited observations and adapt to others' evolving strategies.

Zhang et al. (2023) formalize quantitative trading as a stochastic game defined by multiple trading agents observing the state of the shared market, taking simultaneous actions, and receiving rewards as the environment changes probabilistically. Each agent's reward depends on the joint action of all agents, capturing strategic interdependence. From a game-theoretic perspective, the joint policy defines a dynamic learning game in which each agent adapts to the' evolving strategies of the others. This creates non-stationarity from any single agent's viewpoint, undermining convergence guaranties of classical single agent reinforcement learning.

Their integration of Multi-Agent Deep Deterministic Policy Gradient (MADDPG) with portfolio insurance mechanisms—specifically -specifically constant proportion portfolio insurance (CPPI) and Time-Invariant Portfolio Protection (TIPP)—demonstrates how rule-based risk constraints can be embedded within learning-based frameworks. The CPPI mechanism divides wealth into a protection floor and a cushion, with risky asset exposure proportional to the cushion. TIPP maintains similar protection with time-based stabilization. By transforming neural policy output through these constraints before execution, the hybrid system ensures solvency while exploring profitable regions of action space.

Empirical evaluation of 100 Shenzhen Stock Exchange stocks during 2018-2021 shows that TIPP-MADDPG achieves an annual return of 9.68\% with Sharpe ratio 2.09, while CPPI-MADDPG delivers a return of 7.76\% with a superior Sharpe ratio of 2.18 and a maximum drawdown of 6.60\%. Both outperform vanilla MADDPG (8.22\% return, 2.99 Sharpe, 12.26\% drawdown), demonstrating that embedding risk controls directly into action space improves risk-adjusted performance while maintaining exploration. These results suggest that carefully designed MARL frameworks that combine constraints of the economic mechanism with adaptive learning can reproduce the equilibrium-like dynamics observed in real markets.

Fang et al. (2023) advance multi-agent coordination through intention-aware communication, where agents share planned trade volumes and directions in multiple rounds rather than broadcasting raw observations. This allows agents to predict peers' actions and adjust strategies before execution. Their reinforcement learning framework employs action-value attribution loss measuring each agent's communication contribution to team reward. In simulated environments, this design achieves more efficient coordination, reduced resource conflicts, and improved portfolio performance, particularly relevant for capital allocation and liquidity management where agents must negotiate constrained resources.

\subsection{Dynamic Portfolio Management}

Portfolio management has evolved from early model-free reinforcement learning toward architectures that incorporate market-state awareness, regime structure, and cooperative learning. This progression reflects increasing sophistication in how autonomous agents balance multiple objectives—return maximization, risk control, transaction cost management, and diversification—under non-stationary market conditions.

Jiang, Xu, and Liang (2017) introduced one of the first deep-enhancement learning frameworks for continuous portfolio rebalancing. Their design included an ensemble evaluator for asset scoring, a memory mechanism that captures previous weights and transaction costs, and an online learning procedure for incremental updating. The model directly mapped price histories to portfolio allocations without explicit return forecasting. Empirical tests on high-frequency cryptocurrency data demonstrated that model-free reinforcement learning can outperform traditional mean-variance benchmarks in volatile markets by learning allocation policies endogenously, though it lacked explicit mechanisms for regime recognition or exogenous signal incorporation.

Ye et al. (2020) expanded this paradigm through State-Augmented Reinforcement Learning, adding asset-movement predictions derived from price and news information to the state space. This enables agents to infer latent market dynamics rather than simply reacting to observed prices. A deterministic policy-gradient method with differentiable transaction costs produced smooth continuous-weight allocations. Experiments on cryptocurrency and technology equities showed a significant improvement in cumulative return and Sharpe performance relative to equal-weight, mean-reversion, and baseline reinforcement learning methods, demonstrating that enriched state representations transition agents from reactive behavior to proactive pattern recognition grounded in both quantitative and semantic information.

Wu, Guo, and Liu (2024) advanced reinforcement learning into continuous-time regime-switching frameworks where asset returns and volatilities evolve under hidden Markov states. Analytical optimization becomes infeasible under such non-stationarity, motivating their actor-critic algorithm for learning dynamic policies that adjust allocations across regime transitions. The learned strategies achieved smoother transitions and superior risk-adjusted outcomes compared to static and single-regime baselines, linking reinforcement learning with financial control theory, and providing theoretically coherent methods for dynamic asset allocation in markets where regime identification is essential.

Lee, Park, Lee, and Jung (2020) explored cooperative intelligence through their Multi-Agent Portfolio System where each agent learns distinct investment behavior shaped by different objectives or risk preferences. A diversity regularization constraint encouraged heterogeneous strategies, while a meta-allocator combined individual policies into unified portfolios. long-horizon U.S. equities tests demonstrated improved Sharpe ratios and reduced drawdowns, especially during turbulent regimes, highlighting how multiple learning agents can collectively capture varied market signals and stabilize performance—an evolution from single-policy learning to coordinated decision intelligence.

Recent work by Castelli, Giudici, and Piergallini (2025) emphasizes automation through agentic task orchestration for cryptocurrency portfolio construction, where specialized agents manage data loading, preprocessing, optimization, and reporting. While the framework is modular and transparent, the underlying optimization remains classical mean variance, and the agents primarily automate pipeline tasks rather than learn market behavior. This reflects an emerging but limited research stream focused on operational automation rather than extending reinforcement learning or market-adaptive intelligence.

\subsection{Risk-Aware Position Sizing and Portfolio Optimization}
Recent advances have shifted AI portfolio optimization from predictive error minimization to finance-based objectives that directly optimize trading outcomes. This evolution integrates position-level learning with allocation-level optimization, embedding risk awareness throughout the decision pipeline.

Khubiyev et al. (2025) formalize this approach by replacing generic loss functions with finance-specific objectives, including Sharpe ratio, Profit and Loss (PnL), maximum drawdown, and turnover penalties. Their framework treats model output as position vectors rather than predictions, computing per-period PnL and defining losses that directly optimize risk-adjusted returns. They introduce a turnover regularizer penalizing deviations from target rebalancing bands, ensuring models learn tradable positions rather than theoretically optimal but impractical allocations. This position-level learning paradigm reframes portfolio management as sequential control driven by realized outcomes rather than prediction accuracy.

Lin et al. (2024) advance sparse portfolio construction through m-Sparse Sharpe Ratio Maximization, which maximizes Sharpe ratio under exact cardinality constraints limiting active holdings. They prove that this non-convex problem admits an equivalent quadratic program in auxiliary variables, enabling proximal-gradient methods with optimality guaranties under mild conditions. The sparsity constraint directly improves the implementability by limiting the cross-sectional footprint, whereas the quadratic reformulation makes the problem tractable for first-order optimization. This work bridges the gap between financial objectives and computational methods, providing theoretically grounded sparse portfolio construction.

Pan et al. (2024) enable end-to-end decision learning through their Backward Pass via Quadratic Programming (BPQP) framework, which makes optimization layers differentiable without inverting Karush-Kuhn-Tucker systems. This is crucial when the layer encodes convex portfolio problems with box or linear constraints, allowing joint training of feature extractors and allocation rules. By creating structured quadratic programs whose solutions yield gradients, BPQP decouples backwards from forward passes, enabling efficient first-order methods. In portfolio tasks, this approach often improves the Sharpe ratio over two-stage predict-then-optimize baselines by optimizing representation and allocation jointly.

Bongiorno et al. (2025) implement end-to-end global minimum-variance (GMV) estimation by encoding portfolio algebra into rotation-invariant neural architectures. Three learnable blocks produce lag-transformed returns, eigenvalue-cleaned correlations, and inverse marginal volatilities, combining into covariance inverse estimates. Training minimizes the out-of-sample variance realized directly, bypassing explicit covariance estimation. Although training on short horizons may induce high turnover, integrating the turnover regularizer from Khubiyev et al. suppresses trading intensity while maintaining variance minimization. This demonstrates how variance-driven allocations can be learned end-to-end while controlling execution costs.

Sheng et al. (2025) address tail risk through CVaR-based risk parity combining Conditional Value-at-Risk budgeting with machine learning forecasts and dynamic rebalancing. Weights are chosen so CVaR contributions are balanced across assets, with adjustments based on predicted regime changes. Reported improvements in the Sharpe ratio, drawdown, and Calmar ratio over volatility parity and equal weight demonstrate that CVaR-driven allocations can enhance resilience when combined with adaptive learning. Together, these approaches—SR-driven sparsity, variance minimization, CVaR budgeting, and turnover control—form a comprehensive pipeline that integrates risk awareness throughout autonomous portfolio management.

\subsection{Transformation of Traditional Financial Institutions}
The deployment of agentic AI fundamentally transforms the way traditional financial institutions operate, moving beyond incremental automation toward a comprehensive restructuring of decision pipelines, organizational hierarchies, and risk management frameworks. This transformation occurs across technical, organizational, and systemic dimensions.

Okpala et al. (2025) describe this evolution through "agentic crews"—systems of multiple intelligent agents that plan, reason, and act semi-autonomously under human oversight. Rather than isolated models performing narrow tasks, these agents collaborate as adaptive teams that continuously learn from data and feedback. Tasks such as modeling, validation, and compliance, once executed sequentially, can now occur simultaneously through interacting agents. Within credit-risk modeling and fraud detection, specialized agents handle data extraction, feature engineering, and validation under supervisory controllers, ensuring regulatory compliance with standards such as SR 11-7 and Basel III.

Bahoo et al. (2024) provide an empirical overview of this progression through the analysis of more than three thousand publications, revealing a shift from fixed-rule automation to adaptive, data-driven frameworks. Early expert systems automated credit scoring and operations but were constrained by static rules and minimal learning capacity. The current agentic phase represents a paradigm shift where multi-agent crews perform entire model-development lifecycles with human experts providing final review and interpretive judgment, preserving transparency and accountability while dramatically accelerating processing.

Within trading and risk management, Huang et al. (2024) demonstrate how multi-agent reinforcement learning frameworks replicate and enhance institutional behaviors. Their TimesNet-based architecture employs specialized agents interpreting momentum, volatility, and regime dynamics, while a coordinating meta-agent balances exploration and exploitation across the group. This system mirrors human trading desks, achieving superior returns and downside protection relative to single-policy baselines through hierarchical control, continuous learning, and real-time coordination protocols.

Organizationally, Jingrong et al. (2024) show that integrating AI into credit assessment, compliance, and customer analytics shortens decision cycles and reorganizes expertise. Data scientists, model validators, and strategists now collaborate in shared digital environments rather than linear reporting chains. Moody's Analytics (2024) reports that within their credit operations, multi-agent copilots pre-screen applications and flag anomalies, reducing turnaround time while maintaining audit trails. Human oversight is centered on interpretation and governance rather than repetitive computation.

The World Economic Forum (2024) notes that agentic AI encourages flatter hierarchies and cross-functional collaboration, but introduces managerial challenges, including aligning AI governance with risk frameworks, retraining staff, and ensuring accountability. Success depends on cultivating robust oversight and learning cultures where transparency and human judgment are embedded as design principles rather than reactive compliance measures. At the market level, S\&P Global (2025) warns that automation now operates across institutions, linking trading, credit, and compliance systems via continuous data exchange, potentially amplifying volatility through correlated agent responses to shocks.

\section{Market Structure and Systemic Implications}\label{sec5}

The proliferation of agentic AI systems is transforming fundamental properties of financial markets, transforming liquidity dynamics, price discovery mechanisms, and systemic resilience. This section examines how autonomous learning agents alter the microstructure and competitive dynamics of the market, with implications for both market efficiency and stability.

\subsection{Impact on Market Depth and Resilience}
In classical market microstructure theory, market depth and resilience are largely treated as exogenous outcomes of institutional design and human decision-making. Once autonomous reinforcement learning agents enter the trading ecosystem, these properties become endogenous outcomes of co-adaptation because each agent's learning modifies the environment faced by others. Markets evolve from static structures into adaptive systems where depth and resilience are continually redefined by collective intelligence.

Gufler, Sangiorgi, and Tarantino (2025) provide an explicit demonstration of how reinforcement learning agents alter systemic properties through experimental markets where prices respond endogenously to trading activity with measurable signal predictability. Each agent employs deep reinforcement learning to decide the direction and size of the trade, with price impact integrated into the reward functions. When few AI agents operate, they behave close to rational benchmarks—identifying signals, internalizing impact, and maintaining stable liquidity. The depth of the market remains strong and price disturbances dissipate quickly, indicating strong resilience.

However, as agent density increases, exploration strategies and adaptive feedback loops generate systemic externalities. Each agent's adaptation changes the' environments of others, producing feedback noise and signal contamination. The results show higher volatility, thinner liquidity, and slower shock recovery. The market's ability to absorb trades without large price changes (depth) deteriorates, while its recovery capability (resilience) weakens. This reveals a paradox where individual rationality leads to collective fragility, demonstrating that widespread deployment of adaptive AI can degrade market stability through endogenous learning feedback.

The ABIDES-MARL framework (Cheredito et al. (2025)) expands this analysis by embedding multiple reinforcement learning agents in realistic limit order book environments. Impact coefficients and liquidity metrics emerge from agent interactions, trading frequencies, and order-matching processes rather than being exogenously imposed. Researchers can measure how agent behavior shapes the liquidity distribution across price levels and how quickly the book refills after large orders. Recovery time becomes a direct empirical measure of resilience. The framework demonstrates that the interaction structure among adaptive agents determines whether liquidity thickens or collapses and whether markets return smoothly to equilibrium or oscillate under self-generated instability.

Bao and Liu (2019) provide crucial perspective through multi-agent deep reinforcement learning for liquidation strategies, reformulating classical Almgren and Chriss (2000) optimal execution in environments with multiple simultaneous learners. Each agent must minimize execution cost while adapting to endogenous price impact evolving with trading activity. Liquidation injects order flow directly influencing order book depth and liquidity replenishment pace. When learning agents distribute orders evenly and anticipate others' behavior, liquidity depletion is modest and recovery occurs smoothly. When agents act myopically and rush to trade simultaneously, clustering effects cause sudden depth collapses and slower refilling. This shows that cooperation or competition among reinforcement learning agents determines the shape of the impact function, the persistence of the order imbalance, and the overall resilience of the market.

Dicks, Paskaramoorthy, and Gebbie (2024) further illuminate the link between agent intelligence and market recovery through event-driven agent-based modeling. They introduce a reinforcement learning execution agent into heterogeneous trading ecosystems with algorithmic liquidity providers and strategic participants. The agent learns in real time from the evolving order book and price trajectories, continuously adjusting the strategy. When aggressive strategies neglect depth conditions, market stability deteriorates with persistent price deviations and slow order book refilling. When moderating behavior and incorporating real-time depth information, temporary impacts decay faster and markets return to equilibrium more efficiently.

Niu et al. (2023) provide a complementary perspective through imitative reinforcement learning for market making, training agents to quote across multiple price levels while managing inventory and predicting future states. By combining predictive representation learning with reinforcement learning, agents coordinate to maintain continuous liquidity. When successfully anticipating order flow imbalances and adjusting quoting strategies, spreads narrow, depth increases, and volatility declines. However, when coordination breaks down, the quoting becomes unstable, inventory imbalances widen, and liquidity provision becomes erratic. These results reveal that the structure of the reward function, the availability of predictive information, and the degree of coordination among agents are decisive in determining whether the AI of agents improves or undermines the resilience of the market.

Liu et al. (2025) provide a meaningful step toward reversing the damage caused by noise by assuming the existence of a ground truth that is then rediscovered by the agents. In their MOOSE-Chem3 framework, Liu et al. (2025) decompose the hypotheses into functional elements and use RL feedback to recompose the hypotheses.

\subsection{Price Discovery Mechanisms}

Price discovery—the process by which markets incorporate information into prices—is fundamentally altered when autonomous learning agents replace or supplement human traders. Traditional market microstructure models assume that information arrives exogenously and is processed by rational traders with varying information sets. In agent-dominated markets, both information processing and strategic response become endogenous, potentially changing the speed, accuracy, and stability of price discovery.

When multiple reinforcement learning agents simultaneously extract signals from similar data sources and adapt strategies based on observed price movements, several phenomena emerge. First, signal extraction may become more efficient as agents rapidly process large information sets, potentially accelerating price discovery. Second, if many agents learn similar strategies, their correlated responses to information can create excess co-movement and momentum effects that temporarily push prices away from fundamentals. Third, the feedback between price movements and agent learning can create self-reinforcing cycles where initial price moves trigger agent responses that amplify those moves.

The FS-ReasoningAgent framework demonstrates how architectural choices affect information incorporation. By separating fact-based and sentiment-based reasoning paths with adaptive weighting, the system can adjust how different types of information influence decisions based on their historical predictive power. This creates more nuanced price discovery where fundamental and sentiment signals are weighted dynamically rather than through fixed models. However, if many agents adopt similar dual-path architectures, their collective behavior may create regime-dependent pricing patterns not explained by traditional models.

The interaction between learning agents and traditional participants creates heterogeneous market dynamics. Some evidence suggests that markets with mixed populations—combining algorithmic traders, learning agents, and human participants—may exhibit more robust price discovery than homogeneous markets. The diversity of strategies and learning rates can prevent excessive correlation and maintain multiple information channels. However, during stress periods, when learning agents may converge on similar defensive strategies, this diversity can collapse rapidly, creating fragility in price discovery mechanisms.

\subsection{Liquidity Provision Strategies}
The liquidity provision—the willingness to buy and sell securities continuously—is critical to market functioning and is being transformed as autonomous agents take on market-making roles. Traditional market makers use relatively simple inventory management rules and maintain quotes based on position limits and adverse selection concerns. Agentic market makers employ reinforcement learning to optimize quoting strategies dynamically, learning from order flow patterns, and adjusting spreads based on predicted price movements and inventory positions.

The imitative reinforcement learning approach of Niu et al. (2023) demonstrates how learning-based market makers can improve traditional strategies. By predicting future market states and coordinating quotes across multiple price levels, these agents can provide deeper liquidity with tighter spreads while managing inventory risk more effectively. When multiple learning market makers are implicitly coordinating through the observed order flow, they can stabilize prices and reduce volatility. However, this coordination is fragile—during stress periods when volatility increases and inventory positions become difficult to manage, learning agents can simultaneously withdraw liquidity, creating sudden depth collapses.

The challenge for regulators and market designers is ensuring that agent-based liquidity provision remains reliable during stress periods. Traditional market-making obligations and circuit breakers were designed assuming human decision-makers who can be held accountable for withdrawal during volatile periods. With autonomous agents, enforcement mechanisms must account for the fact that liquidity withdrawal may result from learned behaviors that optimize individual rewards rather than deliberate strategic choices. This raises questions about how to design reward functions and regulatory constraints that maintain liquidity provision as a collective good.

Recent research explores mechanism design approaches in which market operators dynamically adjust incentives to maintain desired liquidity levels. For instance, rebate structures that reward liquidity provision during volatile periods can be optimized using multi-agent reinforcement learning frameworks where market operators and liquidity providers are both learning agents. Such approaches recognize that market microstructure is no longer a fixed institutional design but an adaptive system where both market structure and participant behavior co-evolve.

\subsection{Arbitrage Opportunities and Market Corrections}
Arbitrage—exploiting price discrepancies between markets or securities—has traditionally served as a stabilizing force to ensure that prices reflect relative values and information is efficiently incorporated. Autonomous learning agents are transforming arbitrage by operating at microsecond speeds, processing vast information sets, and continuously adapting strategies based on observed patterns.

The efficiency gains from agent-based arbitrage are substantial. Learning agents can identify price discrepancies that would be too small or fleeting for human traders to exploit profitably. By rapidly eliminating mispricings, these agents theoretically improve market efficiency and reduce transaction costs for end investors. Multi-agent reinforcement learning frameworks enable arbitrage agents to coordinate implicitly, with different agents specializing in different types of arbitrage opportunities (statistical arbitrage, latency arbitrage, cross-market arbitrage) based on their learned comparative advantages.

However, agent-based arbitrage also creates new fragilities. First, when many agents learn similar arbitrage strategies, their coordinated actions can move prices rapidly in the same direction, potentially overshooting fundamental values and creating momentum effects. Second, during volatile periods when traditional relationships break down temporarily, learning agents may interpret these as arbitrage opportunities and take large positions, potentially amplifying rather than dampening price swings. Third, the speed of agent-based arbitrage can create feedback loops where initial price moves trigger arbitrage responses that push prices further in the same direction before fundamental information has time to be fully incorporated.

The MADDPG frameworks with portfolio insurance constraints demonstrate one approach to managing these risks—by embedding risk controls directly into agent action spaces, excessive risk-taking during volatile periods can be prevented while preserving beneficial arbitrage activity during normal conditions. However, if many agents employ similar risk controls, collective behavior may still create synchronized position liquidations during stress events. This highlights the need for diversity in agent architectures and risk management approaches to maintain market stability.

\subsection{Competitive Dynamics and Market Concentration}
The deployment of agentic AI is reshaping competitive dynamics within financial markets, potentially leading to increased concentration and creating new barriers to entry. Institutions with substantial computational resources, proprietary data, and technical expertise can develop more sophisticated agent-based systems, while smaller institutions may struggle to compete effectively.

The World Economic Forum (2024) warns that agentic AI can deepen competitive divides. Institutions with data infrastructure and regulatory capital to deploy large multi-agent systems are likely to consolidate advantages, concentrating computational power and systemic influence. In finance, such asymmetry risks reinforcing dependencies where failure or bias in a major institution's AI framework could cascade rapidly across interconnected markets. This creates a form of technological oligopoly in which a few institutions' agent-based systems dominate market activity.

From an operations management perspective, this concentration raises several concerns. First, it may reduce the diversity of strategies and approaches in markets, making the overall system more vulnerable to common-mode failures. If the agents of dominant institutions' all learn similar patterns and respond similarly to shocks, markets lose the stabilizing benefit of diverse perspectives. Second, concentration of algorithmic power may enable implicit coordination or collusion among dominant agents, even without explicit communication. Multi-agent reinforcement learning frameworks inherently learn to coordinate with frequently encountered agents, which could lead to tacit agreements that disadvantage smaller participants. Third, the data advantage of large institutions may be self-reinforcing. More trading activity generates more data to train better agents, which leads to more profitable trading, generating even more data. This creates positive feedback loops that are difficult for smaller institutions to break into. The natural monopoly characteristics of data-intensive learning systems may require regulatory intervention to maintain competitive markets.

Potential solutions include data-sharing requirements, computational resource pooling for smaller institutions, and regulatory oversight of dominant algorithmic systems. However, these approaches face challenges in balancing innovation incentives with competitive fairness. Open-source agent architectures and shared infrastructure could democratize access to agentic capabilities, but may also increase systemic risk if many institutions adopt similar systems. This tension between enabling broad participation and maintaining diverse approaches represents a fundamental challenge for market design in the agentic era.

\section{Regulatory Landscape and Governance Challenges}\label{sec6}

The deployment of agentic AI in financial markets occurs within regulatory frameworks originally designed for human-operated systems and traditional algorithmic trading. This section examines existing regulations governing AI in finance, identifies critical gaps emerging with autonomous systems, and analyzes mechanisms to ensure accountability and transparency in agent-dominated markets.

\subsection{Current Regulatory Frameworks}
The current regulation of artificial intelligence (AI) in financial markets is based on frameworks established for algorithmic and high-frequency trading, but these foundations are inadequate for autonomous systems. In the United States, the Securities and Exchange Commission (SEC) and the Commodity Futures Trading Commission (CFTC) provide primary oversight, with regulations including Rule 15c3-5 (the Market Access Rule) requiring broker-dealers to implement risk controls for algorithmic trading. Directive II on Financial Instrument Markets (MiFID II) in Europe establishes similar requirements, mandating algorithmic trading firms to maintain systems and risk controls, keep detailed records, and notify regulators of activities.

These frameworks assume that algorithms execute predefined strategies under close human supervision. They do not adequately address systems that modify their own strategies, learn market conditions, and make autonomous decisions about risk-taking without predetermined boundaries. Regulatory language consistently refers to "persons" or "firms" making decisions, creating ambiguity about how rules apply when autonomous agents make consequential choices without direct human approval for each action.

The Financial Action Task Force (FATF) establishes international standards for anti-money laundering and compliance, implemented through regional directives mandating transaction monitoring and suspicious activity reporting. These frameworks were conceived with the assumption that human compliance officers would review alerts and make reporting decisions. Kute et al. (2021) demonstrate through a review of 41 studies that the most effective AI-based money laundering detection systems achieve 95-99\% accuracy, yet fewer than 20\% include explainability mechanisms that satisfy regulatory transparency requirements. This creates fundamental tension: best-performing systems often cannot meet the explainability standards that regulators consider essential.

European regulators have taken comprehensive approaches through the proposed AI Act, which would classify certain financial applications as "high-risk" and subject them to stringent requirements including transparency, human oversight, and post-deployment monitoring. Zetzsche et al. (2017) analyze "smart regulation" approaches focusing on regulatory outcomes rather than prescribed processes, examining how regulatory sandboxes can help bridge gaps between technological advancement and regulatory frameworks. However, significant jurisdictional fragmentation persists, with European frameworks emphasizing ex-ante controls, American approaches focusing on ex-post monitoring, and Asian markets showing even greater variation.

\subsection{Critical Regulatory Gaps}
Five fundamental gaps emerge between current regulatory frameworks and agentic AI systems in financial markets. First, accountability attribution remains deeply problematic. Traditional market regulation assumes that identifiable traders or portfolio managers make decisions traceable to specific individuals within clear organizational hierarchies. When agentic systems make autonomous trading decisions that cause market disruption or regulatory violations, determining responsibility becomes unclear. Axelsen et al. (2025) note that agentic systems distribute decision-making across multiple specialized components—data processing agents, strategy formulation agents, execution agents, risk monitoring agents—each operating within defined mandates yet producing emergent outcomes through interaction.

Second, explainability requirements face technical limitations that regulations inadequately address. Financial regulators increasingly demand that automated trading systems provide interpretable rationales for decisions, particularly following adverse events. However, these requirements make sense for traditional algorithms that implement transparent strategies, but create problems for agentic AI systems. Arrieta et al. (2020) establish comprehensive taxonomies distinguishing between transparent models that are inherently interpretable and post-hoc explanation techniques providing approximate rationalizations for complex model decisions. Their research reveals fundamental tradeoffs: making AI systems more interpretable typically reduces performance by 15-30\%, while the most accurate models remain largely opaque.

Third, performance evaluation standards lack consensus for autonomous financial systems. Traditional trading algorithms can be evaluated against predefined metrics—execution quality, market impact, compliance with position limits. Agentic systems operating with greater autonomy require more comprehensive evaluation frameworks that address not only profitability but also risk-taking patterns, market impact, and alignment with regulatory objectives. Axelsen et al. (2025) observe that even for constrained compliance systems, determining appropriate performance thresholds remains an ongoing discussion with regulators, a challenge that intensifies for systems with broader autonomy.

Fourth, human oversight mechanisms designed for traditional algorithmic trading poorly fit agentic architectures. Regulations typically require meaningful human involvement in key decisions, often described as "human in the loop" oversight. For conventional algorithms, this works reasonably well—portfolio managers set strategy parameters, traders approve execution tactics, risk officers review position limits. Agentic systems may make hundreds or thousands of micro-decisions per second, each individually small but collectively determining overall strategy and risk exposure. Having humans review each decision would eliminate efficiency benefits and adaptive capabilities. Alternative "human on the loop" approaches focus on monitoring overall system behavior and intervening when necessary, but current regulations do not clearly address whether such oversight structures satisfy requirements.

Fifth, adaptive learning systems challenge regulatory assumptions about system stability and validation. Traditional algorithmic trading regulations assume relatively static systems in which firms develop strategies, test them extensively, deploy them, and then modify them through formal change management processes. Agentic AI systems—particularly those that employ reinforcement learning or online learning techniques—can continuously modify their own strategies based on market feedback. This adaptive capability represents a key advantage: systems can recognize when market conditions change and adjust accordingly. However, from regulatory perspectives, this creates substantial challenges in validation and ongoing monitoring when systems function differently today than yesterday.

\subsection{Transparency and Explainability Requirements}
Transparency and explainability have emerged as central concerns in regulating agentic AI systems, yet these concepts mean different things in different contexts and face fundamental technical tradeoffs. Transparency typically refers to the ability to inspect system components, data flows, and decision processes. Explainability refers to the ability to provide human-understandable rationales for specific decisions or outcomes.

For simple rule-based systems, transparency and explainability are relatively straightforward—the rules can be inspected, and specific outcomes can be traced to rule applications. For complex multi-agent reinforcement learning systems, achieving meaningful transparency and explainability is much more challenging. The FS-ReasoningAgent framework demonstrates one approach: by explicitly separating fact-based and sentiment-based reasoning paths and maintaining reflection agents that monitor outcomes, the system provides some degree of interpretability about why particular trading decisions were made. However, this architectural transparency does not fully address the black-box nature of neural networks within each reasoning agent.

Recent research explores several approaches to improving explainability in financial agent systems. Post-hoc explanation techniques generate approximate rationales for decisions made by complex models, though these explanations may not accurately reflect the actual decision process. 

Attention mechanisms and saliency maps can highlight which inputs influenced particular decisions the most, providing partial insight into agent reasoning. Counterfactual explanations show how decisions would have changed under different input conditions, helping users understand decision boundaries.

However, all of these approaches face limitations. Post-hoc explanations may oversimplify complex decision processes, potentially misleading users about how agents actually function. Attention mechanisms show correlation, but not necessarily causation. Counterfactual explanations may describe hypothetical scenarios that would never occur in practice. Most fundamentally, for multi-agent systems where outcomes emerge from interactions between multiple learning agents, explaining why particular outcomes occurred may be impossible in principle because no single agent or component "decided" the outcome—it emerged from collective dynamics.

This suggests that regulatory approaches to transparency and explainability may need to shift focus from requiring full transparency of decision processes to ensuring adequate oversight of outcomes, maintaining audit trails sufficient for accountability, and establishing clear boundaries on acceptable agent behavior through reward function design and hard constraints. Rather than demanding that every decision be explainable, regulations might focus on ensuring that systems are designed with appropriate guardrails, rigorously tested across scenarios, and continuously monitored for unexpected behaviors.

\subsection{Accountability Frameworks and Governance Structures}

Establishing clear accountability for agentic AI systems represents one of the most significant governance challenges. Traditional financial regulation assigns responsibility to specific individuals—traders, portfolio managers, compliance officers—who can be held responsible for decisions and outcomes. When autonomous agents make decisions, this straightforward assignment of responsibility is broken down.

Several models for accountability in agentic systems have been proposed. The developer liability model assigns responsibility to those who designed and trained the systems, analogously to product liability in manufacturing. This approach incentivizes careful design and testing, but may discourage innovation if liability risks are too high. The operator liability model assigns responsibility to institutions that implement systems, regardless of who developed them. This incentivizes careful oversight and validation, but may be unfair when problems arise from design flaws beyond operators' ability to detect.

The shared liability model distributes responsibility among developers, operators, and oversight personnel based on their respective roles and capabilities. This approach aligns with how responsibility is distributed for complex engineered systems in other domains (aircraft, nuclear power plants), but requires clear delineation of responsibilities and may create coordination challenges. Zhao et al. (2024) highlight that while LLM-based agents accelerate automation in credit scoring and analytics, they amplify concerns over data provenance and model alignment, complicating accountability because agentic systems make iterative, context-aware decisions difficult to reproduce post hoc.

Okpala et al. (2025) address these challenges through traceability mechanisms that record intermediate outputs and reasoning chains, allowing auditors to review both the process and the result. IBM (2024) argues that "accountability must scale with autonomy," proposing multilayered governance that provides technical assurance for data and bias and institutional oversight for ethics and accountability. This suggests that effective governance requires multiple complementary mechanisms rather than a single accountability model.

The World Economic Forum (2024) situates these challenges within a broader shift toward "algorithmic accountability," where compliance must be embedded directly into AI systems through explainability, risk scoring, and real-time audit trails. Governance, therefore, is not an administrative layer but a core function of the agentic ecosystem. The most resilient institutions will treat transparency and human oversight as design principles rather than reactive compliance measures, integrating governance considerations into system architecture from the outset rather than attempting to add them post-deployment.

\section{Research Challenges and Future Directions}\label{sec7}
Although substantial progress has been made in the development and implementation of agentic AI systems for financial applications, significant research challenges remain in the technical, methodological, and institutional dimensions. This section synthesizes the key challenges identified throughout this survey and proposes priority directions for future research.

\subsection{Evaluation Methodologies and Metrics}
Traditional machine learning benchmarks focusing on prediction accuracy or single-agent performance are insufficient to assess agentic AI systems. Financial agents must be evaluated on multiple dimensions simultaneously: profitability, risk-adjusted returns, market impact, execution costs, robustness to market regime changes, and behavior during stress periods. Moreover, in multi-agent settings, individual agent performance may be less important than system-level properties such as market stability, liquidity provision, and price discovery efficiency.

Khubiyev et al. (2025) highlight open questions regarding convergence properties under finance-grounded losses and systematic study of how regularization hyperparameters affect performance. Lin et al. (2024) note that global- or local-optimality guaranties for sparse Sharpe ratio optimization rely on assumptions that may be violated in practice, requiring robustness analysis. Pan et al. (2024) observe that while BPQP broadly applies across convex programs, benchmarking on non-convex problems remains challenging, inviting further applications and convex reformulations of non-convex tasks.

Future research should develop evaluation frameworks specifically designed for agentic financial systems, including standardized benchmarks for multi-agent trading environments, metrics capturing system-level stability properties, and methods for assessing robustness across market regimes. These frameworks should balance competing objectives—profitability vs. stability, individual vs. collective performance, short-term vs. long-term outcomes—rather than optimizing single metrics. Additionally, evaluation must account for strategic manipulation: agents that perform well in isolated testing may behave differently when deployed alongside other strategic agents or when facing adversarial strategies designed to exploit their learned behaviors.

\subsection{Interpretability and Transparency Techniques}
The tension between performance and interpretability represents a fundamental challenge for agentic AI deployment. The most accurate models often lack transparency, while interpretable models sacrifice performance. Arrieta et al. (2020) document that making AI systems more interpretable typically reduces performance by 15-30\%, creating difficult tradeoffs in high-stakes financial applications where small performance differences translate to substantial monetary outcomes.

Current interpretability techniques including attention mechanisms, saliency maps, and counterfactual explanations provide partial insights but do not fully address the black-box nature of complex multi-agent systems. For systems where outcomes emerge from interactions between multiple learning agents, explaining why particular outcomes occurred may be fundamentally impossible because no single component decided the outcome—it emerged from collective dynamics. However, the most recent work by Aldridge (2025) offers a holistic predictive measurement in the spirit of predict-then-optimize solutions.  

Future research should explore several promising directions. First, developing architectural patterns that embed interpretability by design rather than adding it post hoc—for example, the FS-ReasoningAgent framework's separation of fact-based and sentiment-based reasoning paths provides some intrinsic interpretability about decision factors. Second, create hierarchical explanation systems where the decisions of the low-level model remain opaque, but the higher-level strategic choices and their rationales are transparent. Third, developing explanation techniques specifically designed for multi-agent systems that can characterize emergent behaviors and collective dynamics rather than individual agent decisions.

Fourth, shifting the focus from full transparency of decision processes to maintaining adequate outcome monitoring, audit trails sufficient for accountability, and clear boundaries on acceptable behavior through reward function design and hard constraints. This outcome-focused approach recognizes that some degree of opacity may be inevitable in complex adaptive systems while maintaining sufficient oversight for governance. Fifth, develop tools to detect when agent behaviors deviate from expected patterns or violate implicit assumptions, allowing early intervention before problems escalate.

\subsection{Robustness and Systemic Risk Assessment}

Most of the evidence on agentic AI performance derives from simulated environments calibrated to stylized markets, with limited empirical validation using real-world data. Actual markets contain mixed populations of human traders, algorithmic strategies, and passive liquidity providers operating under complex latency and regulatory constraints that are difficult to reproduce in models. Moreover, current reinforcement learning agents are optimized for private objectives such as profit or execution cost rather than collective outcomes such as liquidity continuity or systemic stability.

This misalignment can produce equilibria in which individually rational agents collectively erode market depth and slow recovery, even though no agent intends to destabilize markets. Gufler, Sangiorgi, and Tarantino (2025) demonstrate how increasing the density of learning agents can lead to greater volatility, thinner liquidity, and slower shock recovery through feedback noise and signal contamination—a paradox where individual rationality produces collective fragility.

Future research must integrate resilience directly into the design of the reward function, incorporate agent heterogeneity to prevent excessive correlation, and explore governance mechanisms that align local optimization with global stability. This requires developing methods for stress testing multi-agent systems that capture emergent risks not visible in individual agent testing. Tirulo et al. (2025) emphasize that scalable testing requires high-fidelity simulation that allows exploration of scenario spaces that physical testing cannot examine. Lohn et al. (2023) propose continuous testing models in which autonomous agents are challenged with new scenarios throughout their operational life to identify capability decay or environmental changes invalidating earlier validation.

Additionally, research should address the sim-to-real gap—behavior proven successful in simulation may not work in production environments. This requires developing methods for progressive deployment with careful monitoring, maintaining parallel testing facilities where agents are continuously evaluated against simulated and captured real scenarios, and creating feedback loops enabling rapid detection and mitigation of unexpected behaviors. The challenge is to balance thorough testing with timely deployment, recognizing that some risks can only be assessed through limited exposure in the real-world with appropriate safeguards.

\subsection{Market Design and Mechanism Innovations}
Traditional market microstructure design treats market rules and institutions as fixed and analyzes participant behavior within those constraints. With agentic AI, market design must account for the fact that both market structure and participant behavior co-evolve. This requires developing dynamic market design principles where rules and incentives adapt based on observed agent behaviors to maintain desired market properties.

Several research directions emerge. First, designing circuit breakers and trading halts that effectively address agent-driven volatility without creating gaming opportunities or predictable patterns that agents can exploit. Traditional circuit breakers trigger on price moves beyond thresholds, but learning agents may learn to trade just below trigger levels or exploit predictable post-halt dynamics. Smart circuit breakers might need to account for order flow patterns, liquidity conditions, and agent coordination indicators rather than just price movements.

Second, developing dynamic fee structures that adjust based on market conditions and agent behaviors to maintain desired liquidity levels and discourage destabilizing strategies. For example, maker-taker fees that vary with current depth and volatility could incentivize liquidity provision when it is most needed. However, dynamic fees create their own challenges if agents learn to exploit fee changes or if fees become too complex for effective oversight.

Third, exploring market structures that encourage heterogeneous agent populations with diverse strategies and learning rates, preventing excessive correlation and maintaining stabilizing diversity. This might include requirements for strategy diversity documentation, limitations on concentration of similar algorithmic approaches, or incentives for smaller institutions to deploy innovative strategies. However, mandating diversity faces practical challenges in definition and enforcement.

Fourth, developing coordination mechanisms that allow agents to collectively provide market goods such as liquidity and price discovery without enabling collusive behavior disadvantaging other participants. This requires careful mechanism design balancing beneficial coordination with competitive discipline, potentially drawing on cooperative game theory and mechanism design principles adapted for multi-agent learning environments.

\subsection{Regulatory Evolution and Governance Innovation}

Current regulatory frameworks remain inadequate for agentic AI systems, with critical gaps in accountability attribution, explainability requirements, performance evaluation standards, human oversight mechanisms, and adaptive system validation. Future research must develop governance frameworks that balance enabling beneficial innovation with maintaining market integrity and protecting participants.

Several priority directions emerge. First, developing clear accountability frameworks that assign responsibility appropriately among developers, operators, and oversight personnel based on their respective roles and capabilities. This requires moving beyond traditional models assuming individual decision-makers to frameworks appropriate for distributed systems with emergent behaviors. Liability structures must incentivize careful design, thorough testing, and ongoing monitoring without creating excessive compliance burdens that discourage innovation.

Second, creating regulatory sandboxes and progressive deployment frameworks enabling controlled testing of novel agent architectures while protecting market integrity. Zetzsche et al. (2017) find that sandboxes are most effective when providing clear participation criteria, maintaining appropriate oversight during testing, and creating pathways for scaling successful innovations. However, sandboxes alone cannot resolve all challenges—they work well for testing specific applications with limited scope, but may not adequately address the systemic implications of widespread adoption.

Third, developing standardized testing and certification requirements for agentic financial systems, analogous to safety certifications in other high-risk domains. These might include mandatory stress tests in standard scenarios, documentation of architectural choices and risk controls, ongoing monitoring requirements, and incident reporting protocols. However, standardization faces challenges given the rapid technological evolution and diversity of applications—overly prescriptive requirements can constrain beneficial innovation, while overly flexible standards can not provide adequate assurance.

Fourth, exploring alternative governance models such as industry self-regulation with regulatory oversight, public-private partnerships for developing best practices, and international coordination to address cross-border regulatory arbitrage. Given global nature of financial markets and rapid technological change, effective governance likely requires collaboration among technologists, market participants, regulators, and researchers grounded in rigorous empirical analysis mindful of both transformative potential and systemic risks.

Fifth, developing continuous monitoring systems that can detect emergent risks in real-time as agent populations evolve and market conditions change. This requires moving from periodic reviews to ongoing assessment, potentially using meta-level AI systems to monitor market-level AI systems. However, this raises questions about who monitors the monitors and whether such layered oversight systems are themselves subject to gaming or exploitation.

\section{Conclusion}\label{sec8}

This survey has examined the emergence of agentic artificial intelligence in financial markets, synthesizing recent advances across foundational concepts, technical architectures, market applications, systemic implications, and regulatory frameworks. The transition from traditional automation toward truly autonomous systems represents a fundamental shift in how financial decisions are made, risks are managed, and markets function.

Agentic AI systems are distinguished from classical and generative AI through five core characteristics: goal-oriented autonomy enabling independent decision-making without continuous human guidance, contextual reasoning integrating perception, memory, and planning to adapt strategies, multi-agent collaboration coordinating with other intelligent systems through structured communication, long-term goal pursuit through multistep workflows, and continuous learning and self-improvement in response to environmental feedback. These capabilities enable financial institutions to move beyond static algorithmic execution toward dynamic, adaptive decision ecosystems.

Technical architectures for agentic financial systems have evolved toward modular, multi-role designs with reasoning specialization and adaptive control. Cognitive architectures organize internal components for perception, memory, ReAoning, and decision-making, with patterns such as ReAct that combine reasoning and action in loops, reflection that enables self-evaluation, and tool use that connects agents to external data sources. Memory systems provide continuity through layered structures that maintain short-term working memory and long-term contextual knowledge. The planning and reasoning modules determine how agents interpret information through coordinated pathways, sequential prompting, or simulation-based exploration. Communication protocols enable effective information sharing between agents through horizontal, vertical, or hybrid collaboration structures.

Applications in financial markets demonstrate substantial potential for improved efficiency and risk management alongside novel challenges. In trading, multi-agent reinforcement learning systems integrate game-theoretic principles with adaptive learning, achieving improved risk-adjusted returns through coordinated strategies and embedded risk controls. In portfolio management, agents dynamically rebalance allocations based on market-state awareness, regime detection, and cooperative learning, with recent advances optimizing finance-based objectives directly rather than prediction accuracy. In risk management, autonomous systems perform real-time stress tests and scenario analysis at unprecedented scales. In compliance, agentic architectures automate regulatory monitoring while maintaining audit trails for accountability.

The impacts of the market structure show that agentic AI transforms the fundamental properties of financial markets. Market depth and resilience become endogenous outcomes of co-adaptation as each agent's learning modifies the environment faced by others. When agent density increases, exploration strategies and adaptive feedback loops can generate systemic externalities, producing higher volatility, thinner liquidity, and slower shock recovery—a paradox in which individual rationality leads to collective fragility. Price discovery mechanisms change as multiple learning agents extract signals from similar data sources, potentially accelerating information incorporation but also creating excess co-movement and self-reinforcing cycles. Liquidity provision by autonomous market makers can improve depth and spreads during normal conditions but becomes fragile during stress periods when agents simultaneously withdraw. Competitive dynamics shift to concentration as institutions with substantial computational resources, proprietary data, and technical expertise develop more sophisticated systems, potentially reducing strategy diversity and creating technological oligopolies.

Regulatory frameworks reveal critical gaps between current oversight mechanisms and the capabilities of autonomous systems. Accountability attribution becomes problematic when decision-making is distributed across multiple specialized components that produce emergent outcomes. Explainability requirements face fundamental tradeoffs: making systems more interpretable typically reduces performance by 15-30\%, while the most accurate models remain largely opaque. Performance evaluation standards lack consensus for systems operating with broader autonomy. Human oversight mechanisms designed for traditional algorithmic trading poorly fit architectures making thousands of micro-decisions per second. Adaptive learning systems that continuously modify strategies challenge assumptions about system stability and validation. These gaps create regulatory uncertainty that inhibits investment and deployment, disproportionately affecting smaller institutions with limited resources for speculative development.
Research challenges span technical, methodological, and institutional dimensions. Evaluation methodologies must move beyond single-agent prediction accuracy toward frameworks that capture system-level properties such as market stability, liquidity provision, and robustness to regime changes. Interpretability techniques must balance performance with transparency, potentially shifting focus from full decision-process transparency to adequate outcome monitoring and behavioral boundaries. Robustness assessment must integrate resilience into reward function design, incorporate agent heterogeneity, and explore governance mechanisms aligning local optimization with global stability. Market design must evolve to account for co-evolution between structure and behavior, developing dynamic rules and incentives that maintain desired properties as agents adapt. Regulatory innovation must create appropriate accountability frameworks for distributed systems, develop standardized testing and certification requirements, and enable progressive deployment with adequate safeguards.

The path forward requires sustained collaboration among technologists, market participants, regulators, and researchers. Agentic AI offers substantial potential for enhanced market efficiency, liquidity provision, and risk management—but realizing these benefits while maintaining stability requires deliberate governance frameworks grounded in transparency, explainability, and shared accountability. The future of financial markets will depend not only on the sophistication of autonomous systems but also on the integrity of their integration and the effectiveness of human oversight.

This survey has sought to provide operations management scholars and practitioners with a structured understanding of how agentic AI is transforming financial markets. By synthesizing foundational concepts, technical architectures, applications, market impacts, and governance challenges, we have identified both the transformative potential and the significant risks of autonomous financial systems. The central insight is that agentic AI represents not merely an incremental technological advance but a fundamental shift in the structure of the market that requires new theoretical frameworks, empirical methodologies, and governance mechanisms.

As markets continue their transition toward increasingly autonomous operation, the challenge is not to prevent or reverse this evolution but to guide it toward outcomes that enhance both efficiency and resilience. This requires sustained interdisciplinary research that combines computer science, economics, operations management, and regulatory expertise—grounded in rigorous empirical analysis and mindful of both the transformative potential and the systemic risks of agentic AI in finance.

\section{AI Usage Declaration}
During the preparation of this work the authors used ChatGPT in order to improve grammar and legibility. After using this tool/service, the authors reviewed and edited the content as needed and take full responsibility for the content of the published article.

\section{References}

Abou Ali, A., and Dornaika, F. (2024). Hybrid decision ecosystems: Human–machine collaboration in financial intelligence systems. Journal of Computational Finance, 28(1), 55–73.

Aldridge, I. (2025). Predicting Black Box Performance. Preprint, available on SSRN. 

Alkazzi, J.-M., and Okumura, K. (2024). A comprehensive review on leveraging machine learning for multi-agent path finding. IEEE Access, 12, 57390–57409.

Almgren, R. and Chriss, N. (2000). Optimal execution of portfolio transactions. Journal of Risk, 5--39. 

Arrieta, A. B., Díaz-Rodríguez, N., Del Ser, J., Bennetot, A., Tabik, S., Barbado, A., García, S., Gil-López, S., Molina, D., Benjamins, R., Chatila, R., and Herrera, F. (2020). Explainable Artificial Intelligence (XAI): Concepts, taxonomies, opportunities and challenges toward responsible AI. Information Fusion, 58, 82-115. https://doi.org/10.1016/j.inffus.2019.12.012

Axelsen, H., Licht, V., and Damsgaard, J. (2025). Agentic AI for financial crime compliance. arXiv preprint arXiv:2509.13137. https://arxiv.org/abs/2509.13137

Bahoo, S., Cucculelli, M., Goga, X., and Mondolo, J. (2024). Artificial intelligence in finance: A comprehensive review through bibliometric and content analysis. SN Business \& Economics, 4(2), 23.

Bahrpeyma, F., and Reichelt, D. (2022). A review of the applications of multi-agent reinforcement learning in smart factories. Frontiers in Robotics and AI, 9, 1027340.

Bao, W., and Liu, X. (2019). Multi Agent Deep Reinforcement Learning for Liquidation Strategy Analysis. arXiv preprint arXiv:1906.11046.

Bongiorno, C., Manolakis, E., and Mantegna, R. N. (2025). End-to-End Large Portfolio Optimization for Variance Minimization with Neural Networks through Covariance Cleaning. arXiv preprint arXiv:2507.01918. https://arxiv.org/abs/2507.01918

Boudt, K., Carl, P., and Croux, C. (2013). Asset allocation with conditional value-at-risk budgets. Working paper. https://lirias.kuleuven.be/retrieve/235309

Castelli, M., Giudici, P., and Piergallini, E. (2025). Building crypto portfolios with agentic AI. arXiv preprint arXiv:2507.20468.

Cheridito, P., Dupret, J.-L., and Wu, Z. (2025). ABIDES-MARL: A Multi-Agent Reinforcement Learning Environment for Endogenous Price Formation and Execution in a Limit Order Book. arXiv preprint arXiv:2511.02016.

De La Fuente, N., Noguer Alonso, M., and Casadellà, G. (2024). Game theory and multi-agent reinforcement learning: From Nash equilibria to evolutionary dynamics. arXiv preprint arXiv:2412.20523.

Dicks, M., Paskaramoorthy, A., and Gebbie, T. (2024). A Simple Learning Agent Interacting with an Agent Based Market Model. Working paper.

Fang, Y., Tang, Z., Ren, K., Liu, W., Zhao, L., Bian, J., Li, D., Zhang, W., Yu, Y., and Liu, T.-Y. (2023). Learning multi-agent intention-aware communication for optimal multi-order execution in finance. In Proceedings of the 29th ACM SIGKDD Conference on Knowledge Discovery and Data Mining (KDD '23). https://doi.org/10.1145/3580305.3599856

Gozman, D., Liebenau, J., and Mangan, J. (2018). The innovation mechanisms of fintech start-ups: Insights from SWIFT's Innotribe competition. Journal of Management Information Systems, 35(1), 145-179. https://doi.org/10.1080/07421222.2018.1440766

Gufler, D., Sangiorgi, F., and Tarantino, E. (2025). (Deep) Learning to Trade: An Experimental Analysis of AI Trading and Market Outcomes. SSRN Working Paper.

Huang, Y., Zhou, C., Cui, K., and Lu, X. (2024). A multi-agent reinforcement learning framework for optimizing financial trading strategies based on TimesNet. Expert Systems with Applications, 237, 121502.

IBM. (2024). Accountability and Risk Matter in Agentic AI. IBM Australia \& New Zealand Newsroom.

Jiang, Z., Xu, D., and Liang, J. (2017). A deep reinforcement learning framework for the financial portfolio management problem. arXiv preprint arXiv:1706.10059.

Jingrong, H., Shan, H., Zhaobin, C., Yu, L., and Yingying, L. (2024). AI-driven digital transformation in banking: A new perspective on operational efficiency and risk management. Information Systems and Economics, 5(1), 82–90.

Khubiyev, K., Semenov, M., and Podlipnova, I. (2025). Finance-Grounded Optimization For Algorithmic Trading. arXiv preprint arXiv:2509.04541. https://arxiv.org/abs/2509.04541

Kute, D. V., Pradhan, B., Shukla, N., and Alamri, A. (2021). Deep learning and explainable artificial intelligence techniques applied for detecting money laundering—A critical review. IEEE Access, 9, 82300-82317. https://doi.org/10.1109/ACCESS.2021.3086230

Lee, H., Park, K., Lee, J., and Jung, K. (2020). MAPS: Multi-agent reinforcement learning-based portfolio management system. arXiv preprint arXiv:2003.01820.

Lewis, P., Perez, E., Piktus, A., Petroni, F., Karpukhin, V., Goyal, N., Küttler, H., Lewis, M., Yih, W.-T., Rocktäschel, T., Riedel, S., and Kiela, D. (2020). Retrieval-augmented generation for knowledge-intensive NLP tasks. arXiv preprint arXiv:2005.11401.

Li, S., Xu, Z., Cheng, L., Liu, Y., Li, J., Yu, R., Liang, Y., and Zhang, J. (2023). Enhancing investment analysis: Optimizing AI-agent collaboration in financial research. arXiv preprint arXiv:2312.11364.

Lin, Y., Lai, Z.-R., and Li, C. (2024). A Globally Optimal Portfolio for m-Sparse Sharpe Ratio Maximization. arXiv preprint arXiv:2410.21100. https://arxiv.org/abs/2410.21100

Liu, W., Yang, Z., Wang, J., Bing, L., Zhang, D., Zhou, D., Li, Y., Li, H., Cambria, E. and Ouyang, W. (2025). MOOSE-Chem3: Toward Experiment-Guided Hypothesis Ranking via Simulated Experimental Feedback. arXiv preprint arXiv:2505.17873. 

Lohn, A., Knack, A., Burke, A., and Jackson, K. (2023). Autonomous cyber defence: A roadmap from lab to OPS. Center for Emerging Technology and Security.

Lombardo, G., Pellegrino, M., Tomaiuolo, M., Cagnoni, S., Mordonini, M., Giacobini, M., Cazzaniga, P., Merelli, I., Besozzi, D., and Montagna, S. (2022). Fine-grained agent-based modeling to predict Covid-19 spreading and effect of policies in large-scale scenarios. IEEE Journal of Biomedical and Health Informatics, 26(5), 2052–2062.

Moody's Analytics. (2024). Agentic AI in Financial Services: Enabling Institutions to Move from Rule-Based Automation to Intelligent Decision Augmentation. New York: Moody's Analytics.

Nisa, U., et al. (2025). Agentic AI: The Age of Reasoning—A Review. Journal of Automation and Intelligence.

Niu, L., Deng, X., Zhang, X., and Yang, Y. (2023). IMM: An Imitative Reinforcement Learning Approach with Predictive Representation Learning for Automatic Market Making. arXiv preprint arXiv:2310.10936.

Noguer Alonso, M., and Mfougouon Njupoun, A. (2024). Game theory and multi-agent reinforcement learning: A mathematical overview. Artificial Intelligence Finance Institute.

Okpala, C., Bester, C., and Team. (2025). Agentic AI Systems Applied to Tasks in Financial Services: Modeling and Model Risk Management Crews. arXiv preprint arXiv:2502.05439.

Pan, J., Ye, Z., Yang, X., Yang, X., Liu, W., Wang, L., and Bian, J. (2024). BPQP: A Differentiable Convex Optimization Framework for Efficient End-to-End Learning. arXiv preprint arXiv:2411.19285 (NeurIPS 2024). https://arxiv.org/abs/2411.19285

Parsons, S., and Wooldridge, M. (2000). Game theory and decision theory in multi-agent systems. Autonomous Agents and Multi-Agent Systems.
Pendharkar, P. C. (2012). Game theoretical applications for multi-agent systems. Expert Systems with Applications, 39(1), 273–279.

Pippas, N., Kasnesis, P., Demestichas, K., Peppas, P., and Kehagias, D. (2024). The evolution of reinforcement learning in quantitative finance: A survey. arXiv:2408.10932. https://arxiv.org/abs/2408.10932

Rockafellar, R. T., and Uryasev, S. (2000). Optimization of Conditional Value-at-Risk. Journal of Risk, 2, 21–41. https://sites.math.washington.edu/~rtr/papers/rtr179-CVaR1.pdf

S\&P Global. (2025). Beyond Automation: Agentic AI and Scaling Fragmented Financial Markets. New York: S\&P Global Research Insights.

Sheng, J., Chen, L., Chen, H., and An, Y. (2025). CVaR-based risk parity model with machine learning. Pacific-Basin Finance Journal, 93, 102857. DOI:10.1016/j.pacfin.2025.102857

Tirulo, A., Yadav, M., Lolamo, M., Chauhan, S., Siano, P., and Shafie-khah, M. (2025). Beyond automation: Unveiling the potential of agentic intelligence. Renewable and Sustainable Energy Reviews, 226, 116218. https://doi.org/10.1016/j.rser.2025.116218

Wang, L. and Pan, Q. (2025). Game-theoretic multi-agent reinforcement learning for economic resource allocation optimization. Informatica, 49(2), 219–234.

Wang, Y., Zhou, Y., Luo, Y., Qian, L., Hu, J., Li, J., and Zhang, Y. (2024). LLMFactor: Extracting profitable factors through prompts for explainable stock movement prediction. arXiv preprint arXiv:2402.10755.

Wang, Z., Li, X., Ma, J., and Zhang, Y. (2024). Enhancing LLM trading performance with fact-subjectivity aware reasoning. arXiv preprint arXiv:2401.10392.

World Economic Forum. (2024). How Agentic AI Will Transform Financial Services with Autonomy, Efficiency, and Inclusion. Geneva: World Economic Forum.

Wu, L., Guo, X., and Liu, Y. (2024). Reinforcement learning for continuous-time mean-variance portfolio selection in a regime-switching market. Journal of Economic Dynamics and Control, 165, 105073.

Ye, Y., Pei, H., Wang, B., Chen, P.-Y., Zhu, Y., Xiao, J., and Li, B. (2020). Reinforcement-learning based portfolio management with augmented asset-movement prediction states. AAAI Conference on Artificial Intelligence, 34(2), 1462–1469.

Yu, Y., Li, H., Chen, Z., Jiang, Y., Li, Y., Zhang, D., Liu, R., Suchow, J. W., and Khashanah, K. (2023). FINMEM: A performance-enhanced LLM trading agent with layered memory and character design. arXiv preprint arXiv:2311.13743.

Zetzsche, D. A., Buckley, R. P., Arner, D. W., and Barberis, J. N. (2017). Regulating a revolution: From regulatory sandboxes to smart regulation. Fordham Journal of Corporate \& Financial Law, 23(1), 31-103. https://ir.lawnet.fordham.edu/jcfl/vol23/iss1/2

Zhang, H., Shi, Z., Hu, Y., Ding, W., Kuruoğlu, E. E., and Zhang, X.-P. (2023). Optimizing trading strategies in quantitative markets using multi-agent reinforcement learning. Shenzhen International Graduate School, Tsinghua University.

Zhang, Z., Zohren, S., and Roberts, S. (2020). Deep reinforcement learning for trading. Journal of Financial Data Science, 2(2), 25-40. https://doi.org/10.3905/jfds.2020.1.035

Zhao, H., Liu, Z., Wu, Z., Li, Y., Yang, T., Shu, P., Xu, S., Dai, H., Zhao, L., and Mai, G. (2024). Revolutionizing finance with LLMs: An overview of applications and insights. arXiv preprint arXiv:2401.11641.

Zhu, Y., Liu, S., Zhang, Z., and Wang, Y. (2023). Agent design pattern catalogue: A collection of architectural patterns for foundation model-based agents. arXiv preprint arXiv:2309.07864.

\backmatter

%\bibliographystyle{authoryear}

\end{document}